\documentclass[12pt]{article}
\usepackage{leqno, amsfonts}
\begin{document}

\def\vp{\varphi}
\newcommand{\stc}{\stackrel}
\newcommand{\bea}{\begin{eqnarray}}
\newcommand{\eea}{\end{eqnarray}}
\newcommand{\ptl}{\partial}
\newcommand{\Dx}{\Delta\xi}
\newcommand{\ep}{\epsilon}
\newcommand{\qarr}{\stackrel{\cal Q}{\longrightarrow}}
\newcommand{\lc}{L^2 (\Sg_3 (t);\ \bf C;\ b^3 (t){\sqrt \omega (t)} d^3\xi)}
\newcommand{\1}{{\bf\hat 1}}
\newcommand{\Oc}{O\left(c^{-2(L+1)}\right)}
%\newcommand{\im}{{\rm i}} % should be substituted
                          %by \newcommand{\im}{{\rm i}}  in LaTex2e
\newcommand{\Sg}{\Sigma}
\newcommand{\bgt}{\bigotimes}
\newcommand{\Sche}{ Schr\"odinger equation\ }
\newcommand{\Schr}{Schr\"odinger representation\ }
\newcommand{\eu}{$ E_{1,3} $}
\newcommand{\euf}{E_{1,3}}
\newcommand{\rif}{V_{1,3}}
\newcommand{\ri}{$V_{1,3}$ }
\newcommand{\ov}{\overline}
\newcommand{\qstc}{\stackrel{\cal Q}{\longrightarrow}}
\newcommand{\defst}{\stackrel{def}{=}}
\newcommand{\h}{\hbar}
\newcommand{\rin}{$V_{1,n}$}
\newcommand{\rinf}{V_{1,n}}
\newcommand{\rn}{$V_n$}
\newcommand{\rnf}{V_n}
\newcommand{\be}{\beta}
\newcommand{\ga}{\gamma}
\newcommand{\beq}{\begin{equation}}
\newcommand{\om}{\omega}

\title
{ Effective Space Quantization  in Friedman-Robertson-Walker Models}

\author{{E.A. Tagirov}\\
{Joint Institute for Nuclear Research, 141980 Dubna, Russia}}

\maketitle

\abstract{
Quantum--mechanical operators corresponding  to canonical
momentum and position of a point--like particle, which follow
from the quantum field theory
in the general Riemannian space-time,  satisfy generally  to a
deformation of the canonical
commutation relations  with $?^{-2}$  as the parameter of  deformation.
For operators of the quasi-Cartesian coordinates in the closed and open
Friedman--Robertson--Walker models, the deformation reproduces the spatial
part of the well--known Snyder formula for quantization of the Minkowsky
space-time. The spatially-flat models are  distinguished apart
by that the deformation is  reduced  exactly to the standard canonical
commutation relations, which correlates remarkably with the fact of
the observed  flatness  of the Universe. Conditions are briefly
discussed for  which the deformation could  have cosmological
manifestations. Key words: quantum mechanics, cosmology, quantized space.
%\PACS {04.60.-m, 04.20.Cv, 98.80.Hw (Quantum Gravity)}
}

\vspace*{5mm}

In paper \cite{TAG1}, quantum mechanics
in general Riemannian space-time was  formulated as quasi-non-re- \-
lativistic
asymptotic of  the one-particle  sector of  the quantum scalar field
theory of a Fock representation. In papers  \cite{TAG2, TAG3},
a construction was studied in an opposite direction
which is more traditional, namely, quantization of geodesic motion.
In the present note,  a particular but very
curious consequence  of the  former, field-theoretical approach is
presented for the cosmological Friedman--Robertson--Walker space-times;
they will be denoted further as $F^{(k)}$, where $k = \ 1,\ 0,\ -1 $
for the closed, spatially--flat and open models, respectively.

Consider quantum theory (QFT) of the  massive linear scalar field
$\vp (x),  \  x\in F^{(k)}$ taking initially  metric  of $ F^{(k)}$
in a more general form:
\begin{eqnarray}
ds^2 &=& c^2 dt^2 - b^2 (t) \tilde{ds^2},  \nonumber\\
\tilde{ds^2} &\defst& \tilde\omega^{(k)}_{ij}(q)  d q^i d q^j,  \quad
 i, j,...= 1, 2, 3. \nonumber
\end{eqnarray}
The spatial coordinates  $ q^i $ and  metric tensor
$\tilde\omega^{(k)}_{ij}$ are taken  dimensionless, that is all lengths
are measured in units of  the scale factor $b(t)$ (that is, in
the case  of a closed model,  the radius
of  the Universe). An explicit form of $\tilde\omega^{(k)}_{ij}$
in  particular  coordinates is  given below. The Fock space  $\Phi$
is defined  through representation   of the  field operator
$\check \vp (x)$  as an expansion by appropriate  solutions   of the
field equation,
which are asymptotic in  $ ?^{-2}$  . The solutions, in turn,
are defined by Cauchy data in a fixed moment $t_0$ of time and in the
case of a time--dependent external field (the metric, in our case),
the Fock representation depends non-trivially  on the moment $t_0$.
The field-theoretical creation  of particles by another moment $t'_0$
is determined by the Bogoliubov transformation of the initial data from
the moment  $t_0$ to  $t'_0$. It is equivalent asymptotically to
solution of the following Schr\"odinger equation:
\bea
 i \hbar  \left(\frac\ptl{\ptl t}+ \frac34 \frac\ptl{\ptl t}\log b(t)\right)
\ \psi(t,\, q) = \hat H (t, q)\, \psi(t,\, q), \label{schr} \\
 \hat H (t, q) =  -\frac{h^2}{2m} b^{-2}(t) \left(\tilde\Delta
- \zeta \tilde R +  \hat O(c^{-2})\right),
\eea
where  $\tilde\Delta $ and $\tilde R$ are  the Laplace--Beltrami operator
and the scalar curvature, respectively,  for the metric
$\tilde\om_{ij} (q)$, the parameter  $\zeta$  determines  a relation
between  $\vp(x)$ and the scalar curvature of $V_{1,3}$ in the initial
scalar field equation;   relativistic  operator corrections denoted
here  and further as  $\hat O(c^{-2})$ should be calculated
from recurrent relations.

The conserved Klein-Gordon sesquilinear form for $\vp (x) $  induces
the standard   inner product in the space   $\Psi_{t_0} $
of  initial data  $\psi (q) \equiv \psi(t_0,\, q)$ :
 \begin{equation}
(\psi_1,\,\psi_2) = \int_{t =t_0} \ov\psi_1 \,\psi_2 \ b^3(t_0)
\sqrt{\tilde\om} d^3 q, \quad  \psi(q) \in \Psi_{t_0},  \label{l2}
\end{equation}
and thus
$ \Psi_{t_0} \sim  L^2 (t_0 =const;\, {\Bbb C};\, \sqrt{\tilde\om} d^3 x)$
It might seem that thus  functions  $\psi (q)$  may  be interpreted
as  amplitudes
of a probability density  to observe  a  real particle in  a small
neighborhood of  the point $q$ of the configuration space $t = t_0$.
However,  one  should keep in view  that  variables  $q^i$  have
to belong   to spectra of  commuting operators, see, e.g.,  \cite{JAU}
Chapter 13,  \cite{TAG2}, Section 6.  In the canonical quantization
of  the  classical mechanics of a particle, the commutativity of
the primary  position observables is plainly postulated. On contrary,
in the field--theoretical approach indicated above,
the   operators of  coordinates and conjugate  momenta
  are defined naturally  in the framework of
QFT \cite{TAG1} and induce  differential  operators   $\hat q^i $ and
$\hat p_i $  which act on  $\Psi_{t_0} $ and are asymptotically Hermitean with
respect the inner product (\ref{l2}).

It is  interesting that, in general, commutators  between
$\hat q^i, \ \hat p_j$  are  not the canonical ones but form {\it a
deformation} of the latter:
 \begin{equation}
[\hat q^i,\, \hat q^j] = \hat O(c^{-4}),
\quad [\hat p_i,\, \hat p_j] = \hat O(c^{-2}),  \label{ccr1}
 \end{equation}
 \begin{equation}
[\hat q^i, \, \hat p_j] = i\hbar \delta^i_j  + \hat O(c^{-2}), \label{ccr2}
 \end{equation}
Of course, in the Minkowski space-time, the deformation
vanishes if Cartesian coordinates are taken as $q^i$ (this is a trivial
particularity of the  result for  $F^{(0)}$ presented below).
It is easy to  perceive  here  an analogy with the well--known fact
that  the Poincare algebra is  a deformation of the Galilei one.

The  quantum--mechanical  Hamilton operator
$ \hat H (t, q)$ is shown in \cite{TAG1}  to be  asymptotically
unitary equivalent to the energy operator induced in
space   $\Psi_{t_0} $  by  the operator of the energy--momentum tensor
of the field  $\check\vp$ acting in the Fock space$ \Phi$. Therefore,
the scheme
presented  supplies the known method of diagonalization of Hamiltonian,
see, e.g., \cite{GRIB}, by  a representation  of the space of states of
the field excitations as a set of spatially localizable states.

  A particular form of  relations (\ref{ccr1}), (\ref{ccr2}) depends
on which  coordinates $q^i$ are chosen  as  {\it observables of  the space
position} of the particle. In quantum  mechanics,  any  information
on a quantum system  contains  always a mixture   of  information on a
classical measuring  system and  the share  of the mixture  depends
on  choice  of the latter system  \cite{ROV}. Considering  systems of
coordinates as formalizations of devices for measurement of position
of the quantum particle, one may expect (and it would be important to
 prove)  that,  among them,  the most  amount of information on the quantum
dynamics of the particle is provided  by  {\it the normal Riemannian
(quasi-Cartesian)  coordinates} $X^i$ in which  the metric of
 $F^{(k)}$ reads  (superscripts are lowered by the Kronecker  symbol
$\delta_{ij}$):
 \begin{equation}
{\tilde {ds}}^2 = dX^i dX_i + \frac{(X_i dX^i)^2}{1 - k X_i X^i},
\end{equation}
where  $ k =  1,\ 0, \ - 1  $
for the closed, spatially flat and open models respectively.
The quasi-Cartesian coordinates  are singled out by that their
coordinate lines are  geodesic ones with respect to the
metric ${\tilde {ds}}^2$  and  therefore determined only
by the geometry of the  space  in the simplest way.
For other systems,  coordinate  lines  have, at least,  one non-zero
proper curvature which is equivalent  to presence of the  field of an
external  force which deviates the line from the geodesic one.

For  $q^i \equiv X^i $,  the general formulae  from  \cite{TAG1} give
the following  expressions for commutators  (\ref{ccr1}), (\ref{ccr2})
in the first non-vanishing  orders:
\begin{equation}
[\hat X^i, \, \hat X^j] = k l_0^2 \hat L^{ij} + \hat O(l_0^3),
\label{sny}\\
\end{equation}
\begin{equation}
[\hat P_i,\, \hat P_j] = k \left(\frac12 -\zeta\right) l_0
\frac{1 + k X_l X^l}{1 - k X_l X^l} \hat L_{ij} + \hat O(l_0^2),
\label{pp}\\
\end{equation}
\bea
[\hat P_i,\, \hat X^j] &=& -i\hbar\Bigl(\delta_i^j   \label{px} \\
 &+& k \left(\frac12 -\zeta\right) l_0
\frac{\delta_i^j + k  X_i X^j}{1 - k X_l X^l} + \hat O(l_0^2) \Bigr),
\nonumber
\eea
where   $ \hat L^{ij} \defst X^i \ptl/\ptl X_j - X^j \ptl/\ptl X_i$ are
generators  of  quasi-regular  representation  of
${\bf o(3)}$--algebra  (the operators of  angular momentum divided by
$i\hbar $),
\begin{equation}
l_0 = \frac{\lambda_C^2}{b^2(t_0)}, \label{l0}
\end{equation}
and   $\lambda_C \defst \hbar/mc $ is  the Compton wave--length.

Relation   (\ref{sny}) is  just  the spatial part  of  the  basic
 formula  of  Snyder's  theory of  the   quantized space-time  \cite{SNY},
the elementary length   of which  is   taken here  to be  dimensionless
(i.e., it   is  measured  as well as  $X^i$ in units of  the cosmological
scale factor  $b(t_0)$); obviously,   $l_0 $    depends  on  the mass  $m$
and  is negligible  except the case  when
\begin{equation}
\lambda_C\ \small{\stc{>}{\sim}}\  b(t_0).\label{lam}
\end{equation}
If the  quasi-Cartesian coordinates are marked off  by some other
units  of  length, both sides of Eq. (\ref{sny}) should be  multiplied
by $b^2 (t_0)$  and then  the elementary length  (\ref{l0})  reads
 \begin{equation}
    l'_0 = \frac{\lambda_C}{b(t_0)} \lambda_C. \label{l'0}
\end{equation}
It should  be noted, however, that, in the present approach, operators
of momentum $\hat p_i $, in general, do not commute, contrary
to the  Snyder's theory and thus a resemblance with the latter do not goes
beyond the formula  (\ref{sny}).

A necessary condition to satisfy asymptotic relation (\ref{sny}) is the
  motion to be quasi-non-relativistic:
\beq
E_{\rm kin}\  < \  mc^2, \quad {\rm i.e.},\quad
      {l'}_0^2 <\!\!\tilde\Delta\!\!> \ {\small \stc{<}{\sim} } \ 1.
\label{l'}
\end{equation}
The subsequent orders in  $c^{-2}$  will give  new restrictions
containing  derivatives of  $b(t)$. For instance, the condition  for
contribution  of the
order  $O(c^{-6})$ to the commutator  (\ref{sny}) were not
greater than the one of the order  $O(c^{-4})$  looks  as follows
(calculations in a more  detail  will be  published elsewhere):
\beq
8 \,\frac{\dot b}c \, l_0\ {\small \stc{<}{\sim}} \ 1. \label{dot}
\end{equation}
More generally, first terms of   $\hat O (c^{-6})$  in  (\ref{sny})
show that  in the arbitrary order  the commutator of the quasi-Cartesian
coordinate has  the following  structure:
\begin{eqnarray}
\lefteqn{[\hat X^i, \, \hat X^j] =}  \\
& & = k \hat {\cal L}^2 (k; b(t_0),\dot b(t_0), \ddot b(t_0),...;
\tilde\Delta; X^l \ptl_l) \cdot \hat L^{ij}, \label{snyx}
\end{eqnarray}
where  $\hat {\cal L}^2$ is an Hermitean operator which is a polynomial
by  $ k, \ \tilde\Delta $ and  $X^i\ptl_i $; obviously, it commutes
with   $\hat L^{ij}$.

For  $\dot b (t) = 0 $,   the asymptotic series for any
$\hat q^i$  can be summed to formal closed expressions  \cite{TAG1},
and,  in this case,  restriction (\ref{l'}) is removed  and  restrictions
of the type (\ref{dot}) do not arise at all.  Note also  the presence
of the  pole  $X^i X_i = 1$      in formulae  (\ref{pp}) and
(\ref{px}) in the case of $ k=1 $. It is  caused by existence
of a focal point for geodesic lines issued  from a point on a
sphere and reflects a topological non-triviality  of the space
in this case.

  The main  conclusion  from  (\ref{snyx}) consists in that, for  $ k=0 $,
that is for the spatially flat model,  the opertors  $\hat X^i$
commute in any order by $c^{-2}$, i.e. exactly. Though, owing
to dependence of the metric on $t$,  the operators
arise initially  from QFT  as  asymptotical sums \cite{TAG1},
one can easily prove \cite{TAG2}  existence of a unitary
transformation  of initial   $\hat X^i$  to the  ordinary form:
$\hat X^i =  X^i \cdot {\bf \hat 1}$.  At the same time,
conjugate  operators  of momentum  are  translations  in virtue  of
the Euclidean symmetry. Therefore, the standard
Heisenberg commutation relations are restored  exactly,
that is with no asymptotical corrections.
Thus, the first principles of quantum theory
single out discretely just the spatially flat models among all
posible FRW ones. This result coincides remarkably  with
the flatness  of the real Universe following from the modern astrophysical
data and commonly accepted  interpretation of them, see, e.g.,
 \cite{KLOP}, Section 3.3.1.

Return, however, to the cases  $k \neq 0$, the more so since
that these models are not excluded from the category of  realistic ones
\cite{STEIN}. Then, in the  asymptotic approximation under consideration,
the space turns out to be quantized up to the moment of time $t_0$,
determined by condition  (\ref{lam}). This going on,  the concept of
the  particle as an individually localizable object loses meaning  owing
to essential non-commutativity of its coordinates. This implies the lack
of the standard Born probability interpretation of $\psi (X)$, as
it was mentioned above.  Apparently, this circumstance  indicates an
essential returning  of the  quantum field $\check\vp (x)$
as the scale factor  $b(t_0)$ becomes  smaller in the direction of
the past and approaches  to the value of $\lambda_C$.  At  that times,
one may expect  qualitative  changes in contribution of the field into
the equation of state of the matter.  This expectation  is supported  by the  paper
\cite{KAD} in which a similar change  of  the  standard representation
of quantum electrodynamics in  the superstrong electromagnetic
field  is discussed, which  determines an effective elementary length.

The space to be essentially quantized in the sense of  the formula
(\ref{sny}), simultaneous fulfillment  of  conditions  (\ref{lam}),
(\ref{l'}) ? (\ref{dot}) is necessary.   Do particles exist
for which  it takes place  for a reasonable dynamics of
 $b(t)$?  One may say at once that such particle can not be  among
the particles  of  GUT  since the latter  "come into the Universe"
as  ultrarelativistic ones and stay   ultrarelativistic much later of
the  moment  $t_0$ determined by condition (\ref{lam}). More interesting
may be the  hypothetical Weak Interacting Massive Particles (WIMP) of
the cold dark matter created directly  by the gravitational field and  not
connected with  the temperature of the GUT matter.  Arguments
\cite{KUZ} exist   that  the Ultra High Energy Cosmic Rays (UHECR) observed
last  decade are generated  by  such particles  with a  mass
 $m {\small \stc{>}{\sim}}  10^{13} {\rm Gev}/c^2 $.

It is worth also to mention the hypothesis \cite{COLD}  on quintessence,
a neutral scalar field  generating an acceleration of  the  expansion of
 Universe in the
(cosmologically) present epoch  $t_0 \sim 10^{10}$ years,  see
references on recent papers
in  \cite{STEIN}.  Now, the field satisfies condition  (\ref{lam})
and   the space is quantized for the quintessence with
the  effective elementary length of the order
of the scale factor.  In a future,  the accelerated  expansion of
the Universe caused by the  negative pressure of quintessence  will relieve
condition  (\ref{lam}) and  formation of localized excitations
(real particles) of the quintessence $\vp_q (x)$  will become  possible.

It should be  emphasized, however, that  the present discussion of
deformation of  canonical commutation relations
(\ref{ccr1}, \ref{ccr2}) is a
concrete  instance of  those  new features of quantum mechanics,
which can  generally occur in different  cases of  interaction of
quantum systems with the intensive  external field.
 A case of the external electromagnetic in this aspect was considered
in the paper  \cite{KAD} mentioned above. Another instance is provided by
commutators  of  the quantum--mechanical
 operators of spin  of the Dirac particle in the external
gravitational field, which are found  in  \cite{TAG4}. They  form
a deformation of the algebra  ${\bf o(3)}$.\\

The author appreciates  Prof. V.G. Kadyshevsky's  attention to
a preliminary  version of the paper.

\end{document}